\documentstyle[psfig]{l-aa}
\def\double {\baselineskip=0.8truecm
             \lineskip=0pt
             \lineskiplimit=0pt}
\def\kms{\,km\,s$^{-1}$}

\def\marc{mag~arcsec$^{-2}$}

\begin{document}

\double
\thesaurus{
	   11.06.2;
	   11.16.1;
	   13.09.1
	   }

\title{1.65~$\rm \bf \mu$m (H-band) surface photometry of galaxies. IV:
observations of 170 galaxies with the Calar Alto 2.2m telescope.
\thanks{based on observations taken at the Calar Alto Observatory, operated 
by the Max-Planck-Institut f\"ur Astronomie (Heidelberg) jointly with the 
Spanish National Commission for Astronomy.}}

\author{A. Boselli \inst{1}
\and G. Gavazzi \inst{2}
\and P. Franzetti \inst{2}
\and D. Pierini \inst{3}
\and M. Scodeggio \inst{4} 
}

\offprints{A. Boselli}

\institute
{Laboratoire d'Astronomie Spatiale, Traverse du Siphon,
F-13376 Marseille Cedex 12, France
\and Universit\`a degli Studi di Milano - Bicocca, P.zza dell'Ateneo Nuovo 1, 20126 Milano, Italy
\and MPI f\"ur Kernphysik, Postfach 103980, D--69117 Heidelberg, Germany
\and European Southern Observatory, Karl-Schwarzschild-Str. 2, D-85748 Garching bei M\"unchen, Germany
}

\date{Received..........; accepted..........}

\maketitle

\markboth{Boselli et al.: NIR surface photometry of Virgo galaxies}{}

\begin{abstract}

We present near-infrared (H band) surface photometry
of 170 galaxies, obtained in 1997 using the Calar Alto 2.2m telescope 
equipped with the NICMOS3 camera MAGIC.
The majority of our targets are selected among bright members of the
Virgo cluster, however galaxies in the A262 and Cancer clusters and in the Coma/A1367 supercluster
are also included.
This data set is aimed at complementing the NIR survey
in the Virgo cluster discussed in Boselli et al. (1997) and
in the Coma Supercluster, presented in Papers I, II and III of this series.
Magnitudes at the optical radius, total magnitudes, isophotal radii and light
concentration indices are derived.\footnote{Tables 1 and 2 are only available in electronic form at
the CDS via anonymous ftp to cdsarc.u-strasbg.fr (130.79.128.5)
or via http://cdsweb.u-strasbg.fr/Abstract.html}

\end{abstract}

\section{Introduction}

This work presents H-band (1.65 $\mu$m) observations of
170 galaxies in the regions of the Virgo cluster, of the A262 and Cancer clusters
and in the Coma Supercluster obtained in 1997 with the Calar Alto 2.2m telescope 
equipped with the NICMOS3 camera MAGIC. This is
an accompanying paper of Paper III (Gavazzi et al. 1999, this issue) where similar 
observations of 558
galaxies obtained with the TIRGO 1.5m telescope are reported.
Since most considerations are in common between the two papers, we preferred to give
them in Paper III to avoid unnecessary duplications.
Only informations which depend on the Calar Alto instrumentation and telescope are given in
full details here.
The paper is organized as follows: Section 2 describes the studied sample, and  
the observations are outlined in Section 3.
Image analysis strategies are discussed in Section 4. 
Preliminary results are given in Section 5 and summarized in Section 6.

\section{Sample definition, observations and data reduction}

The present paper contains the observation of 170 galaxies, primarily selected among late-type  objects belonging to the Virgo cluster.
Out of the 99 observed Virgo (12$^h$ $\le$ RA $\le$ 13$^h$, 0$^o$ $\le$ dec $\le$ 18$^o$) galaxies, 
84 belong to the Virgo Cluster Catalogue (VCC)
of Binggeli et al. (1985) and 15, in the ouskirts of the cluster, were
selected from the CGCG (Zwicky et al. 1961-68). These galaxies have velocities V$<$3000
km/sec, and can thus be considered bona-fide cluster members.
Observations of 73 filler objects are also given, so subdivided: \newline
20 are CGCG galaxies in the A262 cluster 
(1$^h$43$^m$ $\le$ RA $\le$ 2$^h$1$^m$, 
34$^o$31$^{\prime}$ $\le$ dec $\le$ 38$^o$33$^{\prime}$), 
23 are CGCG objects in the Cancer cluster 
(8$^h$11$^m$ $\le$ RA $\le$ 8$^h$25$^m$, 
20$^o$30$^{\prime}$ $\le$ dec $\le$ 23$^o$) and 28 are CGCG
galaxies in the region 11$^h$30$^m$ $\le$ RA $\le$ 13$^h$30$^m$, 
18$^o$ $\le$ dec $\le$ 32$^o$ containing the Coma supercluster, which 
includes the Coma and the Abell 1367 clusters and relatively isolated 
galaxies in the bridge between these two clusters.

By themselves these observations do not form a complete sample in any sense.
However, combined with data published in Paper I (Gavazzi et al. 1996c), II (Gavazzi et al. 1996b) (which were devoted to observations of
disk galaxies), III of this series and in Boselli et al. (1997: B97) (containing mainly measurements of Virgo galaxies taken with the Calar Alto 2.2~m telescope), the present 
survey contains a complete set of NIR observations as follows:
out of the 646 galaxies, of both early and late-types in the CGCG (m$\rm _p \le$~15.7) which are members to the Coma supercluster
($\rm 18^o \le \delta \le 32^o$; $\rm 11.5^h \le \alpha \le 13.5^h$)
according to Gavazzi et al. (1999), i.e. 5000 $<$ V $<$ 8000 \kms, 625 (97 \%) have a NIR image available.
Moreover the survey contains 221 out of 248 (89 \% complete) VCC galaxies brighter than $m_p$=14.0. Thus the giant members of the Virgo cluster (excluding VCC galaxies which are found in the background of the Virgo cluster) are sampled in a quasi-complete
manner. 
A less complete coverage is at $m_p~\leq~16.0$: 277/587 objects were observed (47 \% complete). 
However, we have observed all but one the 88 late-type VCC galaxies selected as part of
the central program of the Infrared Space Observatory (ISO) 
(see B97) brighter than $m_p$=16.0. These are objects lying either within 2 degrees of projected radial distance from M87 or in the corona between 4 and 6 degrees. 
Thus the H band survey contains a complete ($m_p~\leq~16.0$) sample of late-type dwarf members of the Virgo cluster, restricted however to a region smaller than the VCC. \newline

\subsection{Observations}

The observations were carried out in three photometric nights of February 26, 27 and 28,
1997 with 
the Calar Alto 2.2-m telescope. The Cassegrain focus of the 
telescope was equipped with the MAGIC $256 \times 256$ pixel NICMOS3 infrared array 
(Herbst et al. 1993). 
In order to observe galaxies with large apparent sizes, the optical 
configuration of the detector was chosen to give the largest possible field 
of view, i.e. $6.8 \times 6.8$ arcmin$^2$, with a pixel size of 1.61 arcsec. 
The observational technique and the data reduction procedures, here just briefly
summarized, are similar to the one described in B97 and in Paper III.
\newline
The seeing ranged between 2 and 3 arcsec with an average of 2.4 arcsec, as shown in
Fig. 1. These seeing conditions are mostly due to the large pixels in the
selected optical configuration, 
and as such represent a necessary disadvantage, because they
also provide the large field-of-view fundamental for our observations.

\begin{figure}
\psfig{figure=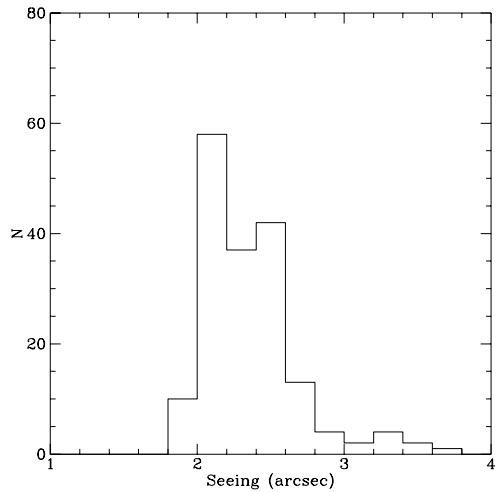,width=10cm,height=10cm}
\caption{The seeing distribution.}
\label{fig.1}
\end{figure}

At H the sky brightness (typically 13.8 mag) varied over the time scale of an observation
by typically 3\% in photometric conditions, by up to
8\% in the worst conditions encountered.
Reaching a brightness limit 8 \marc ~fainter than the sky requires a careful 
subtraction of the sky, necessitating mosaicing techniques. \newline
As in B97 we used three types of mosaic maps, obtained by programming the telescope pointing 
along different patterns. \newline
Galaxies with optical diameter larger than half of the size of the field of view 
of the array were observed using a 
mosaic in which 50\% of the time is devoted to the target of interest and 
50\% to the surrounding sky (``A'' mosaic, Fig 2a in B97). This pattern was obtained 
alternating 8 fields centred on the target with 8 observations of the sky 
chosen along a circular path around the galaxy (off-set by a field of view from the 
centre). The 8 on-target fields were dithered by 10 arcsec in 
order to help the elimination of bad pixels. \newline
Galaxies with optical diameter smaller than half of the size of the field of view 
of the array 
were observed with a mosaic consisting of 9 pointings along a circular path 
and displaced from one-another by 2 arcmin such that the target galaxy is 
always 
in the field (``B'' mosaic; Fig 2b in B97). To avoid saturation
each pointing was split into 32 elementary integrations of 1 sec 
which were added by the on-line MAGIC software. 
There were 7 galaxies with angular sizes larger than the dimension of the 
detector; these were 
mapped using mosaics prepared according to the shape and orientation
of the galaxy in the sky in order to cover the entire surface
of the target. 
In order to get a higher signal-to-noise two observation cycles were secured for the
low surface brightness galaxies. Some galaxies were serendipitously observed 
in the sky frames of other targets. For these objects the number of available frames is generally
$\leq$ 8 (see Table 1), thus their signal to noise is lower than the average value obtained
for pointed galaxies.

\begin{figure*}
\psfig{figure=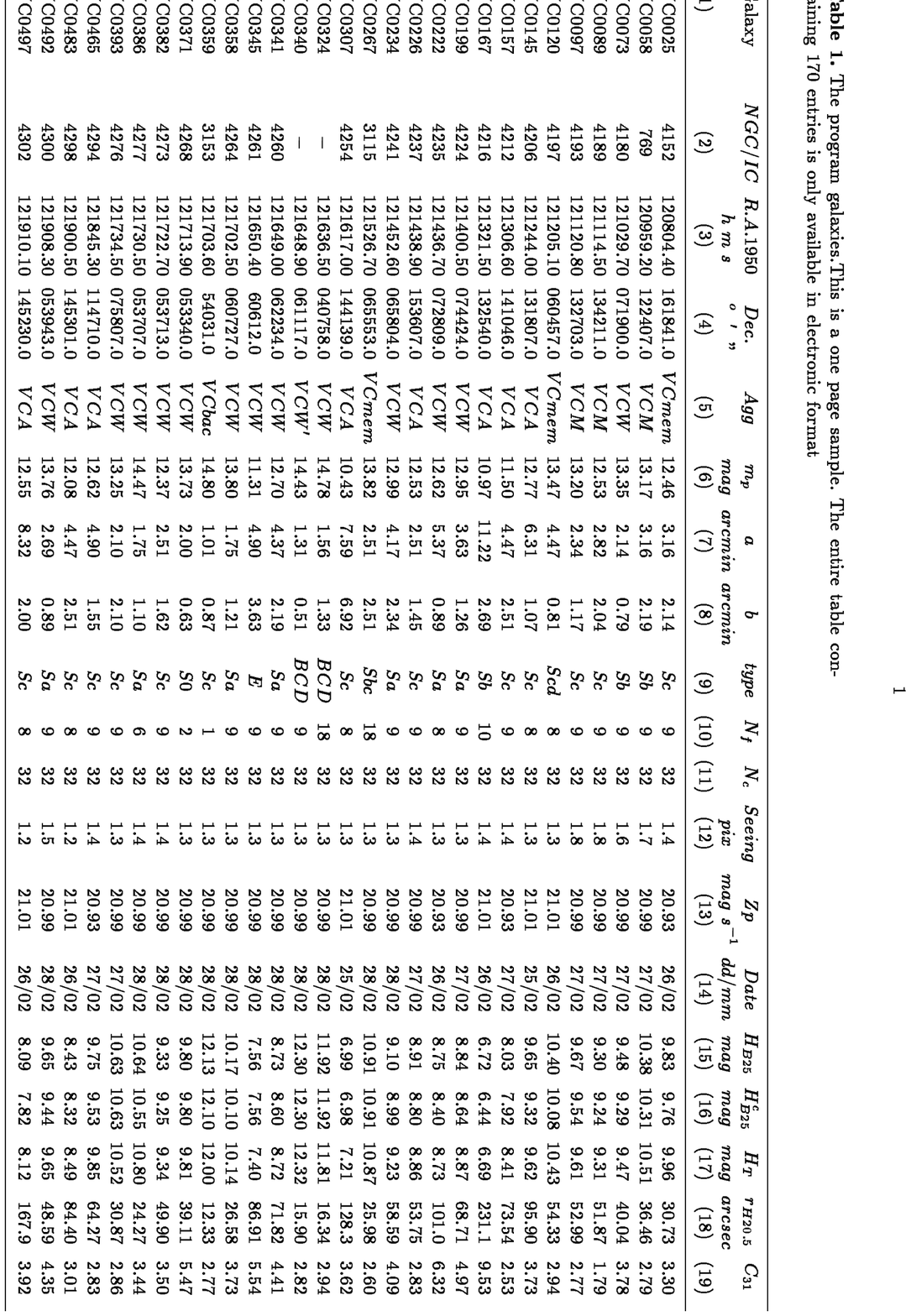,width=18cm,height=23cm}
\end{figure*}

The observations were calibrated and the fluxes transformed into the H
photometric system using standard stars
(Elias et al. 1982), observed hourly throughout the 
night.  Calibration stars were observed with a third mosaic (``C'', Fig 2c in B97). This is 
composed of 5 pointings, starting with the target star near to the centre of 
the array followed by pointings in each of the 4 quadrants of the array.
The observations of the standard stars
were obtained with a defocused telescope to avoid saturation. \newline
The typical uncertainty on the photometric calibration is $\le$ 0.05 mag.

\subsection{Image analysis}

The reduction of two-dimensional IR frames follows procedures identical to those
reported in B97 and in Paper III. These procedures are based on the 
IRAF data reduction package developed by NOAO and on the SAOIMAGE and PROS 
packages developed at the Center for Astrophysics and on STSDAS: \footnote{IRAF is the Image
Analysis and Reduction Facility made available to the astronomical community by
the National Optical Astronomy Observatories, which are operated by AURA, Inc.,
under contract with the U.S. National Science Foundation. STSDAS is distributed
by the Space Telescope Science Institute, which is operated by the Association
of Universities for Research in Astronomy (AURA), Inc., under NASA contract
NAS5--26555.}.
\newline
To remove the detector response, two sets 
of flat-field exposures were obtained on the telescope dome with (lamp-on) and 
without (lamp-off) illumination with a quartz lamp. The response of the 
detector is then contained in the normalized frame 
FF = [(lamp-on) - (lamp-off)] / 
$\langle$ (lamp-on) - (lamp-off) $\rangle$ (per pixel). \newline
Specific reduction strategies were used for the various mosaics, according to
the stability of the sky during the observations. When the sky was stable to
within a few percent during the observation of a galaxy (the large majority of 
the observations), the 8 SKY exposures 
($\rm SKY_i$) were combined using a median filter 
to obtain $\rm \langle SKY \rangle$ for type ``A''
mosaics. For type ``B'' mosaics $\rm \langle SKY 
\rangle$ was obtained by combining the 9 frames containing 
target+SKY with a median filter.  \newline   
The mean counts $\rm \langle c_T \rangle_i$ and $\rm \langle c_{sky} \rangle$ 
were respectively determined for the $\rm i^{th}$ target observation and the 
median sky. Individual ``normalized'' $\rm SKY_i$ frames were then produced 
such 
that $\rm SKY_i$ = $\rm \langle SKY \rangle \times \langle c_T \rangle_i$ / $
\rm \langle c_{sky} \rangle$. 
This removed the time variations of the sky level, but, due to the source
emission, introduced an (additive) 
offset; this was subsequently removed (see below).
Occasionally, when the average response of the detector to the sky changed by
more than 3\% during an observation, significant temporal variations in the spatial
response of the detector to the sky became discernable. Under these 
circumstances, only
the three sky frames closest in time to each target frame were used to
determine the sky.
After sky removal, each target frame 
($\rm T_i$) was processed to obtain a flat-field, sky subtracted, corrected 
frame: $\rm T_{i,corr}$ = [$\rm T_i -  SKY_i$] / FF. 
\newline
Sky-subtracted and flat-fielded frames were then registered using 
field 
stars and combined together with a median filter. This provided a satisfactory
removal of the bad pixels in the final combined image. Tests on the data
showed that the photometry obtained from this use of a median filter was 
identical to that obtained with averaging techniques.
\newline
Star-subtracted frames were produced by a manual ``editing'' of
the contribution from pointlike sources which are clearly not 
associated with the target galaxies. 
\newline
The residual sky background and its rms noise ($\rm \sigma $) (in the 
individual pixels) were determined in each star-subtracted frame in 
concentric object-free annuli around the objects of interest. 
\newline

We checked the quality of the final images on large and small scales. 
On small scales the measured noise was always consistent with the
expected statistical fluctuations in the photon count from the sky background
accumulated over the total integration time. 
The typical pixel to pixel
fluctuations are $\sim$~22\marc~, i.e. 0.05\% of the sky (see Fig. 2).

\begin{figure}
\psfig{figure=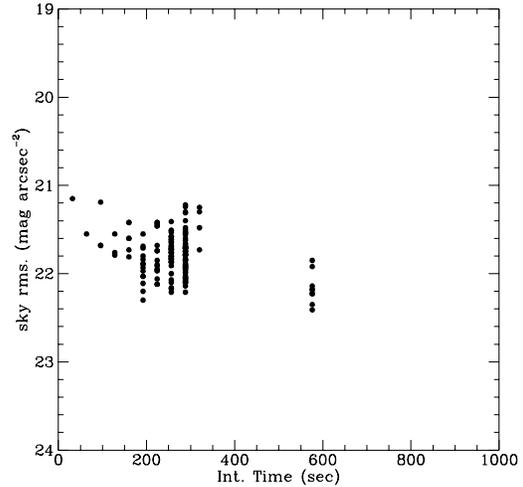,width=10cm,height=10cm}
\caption{The distribution of the sky rms as a function of integration time.}
\label{fig.2}
\end{figure}

\section{Results}

The main results of this paper are given in Table 1 (with structure identical to
Table 1 in Paper III) as follows:
 
Column 1: CGCG (Zwicky et al. 1961-68) or VCC (Binggeli et al. 1985) denomination.  \newline
Column 2: NGC/IC names. \newline
Column 3, 4: adopted (1950) celestial coordinates, measured by us or taken from NED 
\footnote{NASA-IPAC Extragalactic Databasa (NED) is operated by the Jet Propulsion Laboratory,
California Institute of Technology, under contract with NASA},
with few arcsec uncertainty. \newline
Column 5: ``aggregation'' parameter. This parameter defines the membership to a 
group/cluster/supercluster: CSisol, CSpairs, CSgroups indicate members of the Coma Supercluster 
(5000 $<$ V $<$ 8000 \kms); CSforeg means objects in the foreground of the Coma Supercluster 
(V $<$ 5000 \kms) and CSbackg means objects in the background of the Coma Supercluster 
(V $>$ 8000 \kms). Galaxies in the Virgo region are labelled following the membership criteria
given by Binggeli et al. (1993): VCA, VCB, VCM, VCW, VCW', VCSE, VCmem,
are members to the cluster A or B, to the M, W, W' or South-East clouds or are not better specified
members to the Virgo cluster respectively. NOVCC are galaxies taken from the CGCG in the
outskirts of Virgo, but outside the area covered by the VCC. VCback are galaxies in the background
of the Virgo cluster (V$>$3000 km/sec).
Members to the A262 and Cancer clusters are indicated. \newline
Column 6: photographic magnitude as given in the CGCG or in the VCC. \newline
Column 7,8: for CGCG galaxies these are the major and minor optical diameters (a$_{25}$, b$_{25}$) (in arcmin) 
derived as explained in Gavazzi \& Boselli (1996). 
These diameters are consistent with those given in the RC3. For VCC galaxies the diameters are
measured on the du Pont plates at the faintest detectable isophote, as listed in the VCC. \newline
Column 9: morphological type. \newline 
Column 10: number of frames $N_f$ combined to form the final image (depending on the
adopted mosaic). \newline
Column 11: number of elementary observations (coadds) $N_c$. The total integration time (in seconds) is the product of the number of coadds $N_c$ times the number of combined frames $N_f$ 
times the on-chip integration time $t_{int}$ which was set to 1 sec. \newline
Column 12: seeing (in pixels, with 1.61 arcsec per pixel). \newline
Column 13: adopted zero point (mag / sec).  \newline
Column 14: observing date (day-month-1997); \newline
Column 15:  $H_{B25}$ magnitude obtained extrapolating the present photometric 
measurements to the optical diameter along circular apertures as in Gavazzi \& Boselli (1996). \newline
Column 16: $H_{B25}^c$ magnitude computed at the optical diameter (see Column 15) corrected
for galactic and internal extinction following Gavazzi \& Boselli (1996). 
The adopted internal 
extinction correction is $\rm \Delta m=-2.5 \, D\, \log(b/a)$ where D=0.17, as determined 
in Boselli \& Gavazzi (1994).\newline
Column 17: $H_T$ total H magnitude extrapolated to infinity (see Paper V,
Gavazzi et al. 1999a). \newline
Column 18: galaxy observed major ($r_H(20.5)$) radius (in arcsec) determined 
in the elliptical azimuthally--integrated 
profiles as the radii at which the surface
brightness reaches 20.5 H--\marc. Galaxies which require an extrapolation larger than
0.5 mag to reach the $\rm 20.5^{th}$ ~magnitude isophote are labelled -1. \newline
Column 19: the model--independent concentration index $\rm C_{31}$ as defined in de Vaucouleurs (1977) is the ratio between the radii that enclose 75\% and 25\% of the total light $H_T$. \newline

\subsection{The virtual aperture photometry}

The present data were compared with aperture photometry available in the literature 
by integrating the counts in concentric circular rings 
around the galaxy centres to provide curves of growth up to the diameter of 
the reference photometry. This comparison
provided a general check of the intrinsic photometric accuracy of the current 
work. The virtual photometry measurements obtained in this work are compared 
with the aperture photometry available in the literature (200 measurements)
in Fig. 3: on the average we find:

\noindent
$\rm H_{this~work} - H_{literature}$ = -0.013 $\pm$ 0.107 mag.  
\noindent
A conservative estimate of the overall photometric accuracy of our data, including 
systematic errors on the zero point, is thus $\leq $ 0.1 mag.

\begin{figure}
\psfig{figure=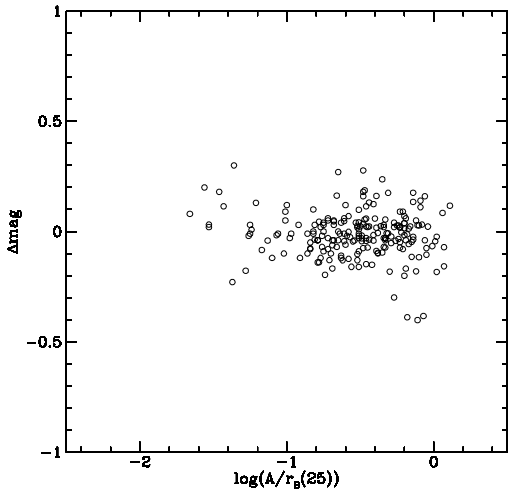,width=10cm,height=10cm}
\caption{The comparison between the present photometric measurements and those available
from the literature as a function of the normalized aperture.}
\label{fig.3}
\end{figure}

The measurements taken through the individual "virtual circular apertures" are given in Table 2
(available only in digital format) as follows: \newline
Column 1: Galaxy denomination in the CGCG (Z) or VCC catalogues. \newline
Column 2: aperture diameter in arcsec.\newline
Column 3: logarithmic ratio of the adopted aperture diameter to the optical a$_{25}$ diameter.\newline
Column 4: integrated H magnitude within the aperture. \newline
 
\begin{table}
\caption{The "virtual aperture photometry". This is a one page sample. The entire table
containing 2646 entries is only available in electronic format.}
\label{Tab2}
\[
\begin{array}{p{0.15\linewidth}cccc}
\hline
\noalign{\smallskip}

Galaxy & Ap. & log Ap/a_{25} & H  \\ 
      & arcsec &  & mag \\
 (1)  & (2) & (3) & (4) \\
\noalign{\smallskip}
\hline
\noalign{\smallskip}

VCC  25  &  19.00 & -1.00 &  10.95 \\ 
VCC  25  &  19.30 &  -.99 &  10.94 \\ 
VCC  25  &  29.00 &  -.82 &  10.55 \\ 
VCC  25  &  38.60 &  -.69 &  10.30 \\ 
VCC  25  &  48.30 &  -.59 &  10.14 \\ 
VCC  25  &  58.00 &  -.51 &  10.06 \\ 
VCC  25  &  67.60 &  -.45 &  10.01 \\ 
VCC  25  &  77.30 &  -.39 &   9.99 \\ 
VCC  25  &  86.90 &  -.34 &   9.97 \\ 
VCC  25  &  96.60 &  -.29 &   9.95 \\ 
VCC  25  & 106.30 &  -.25 &   9.94 \\ 
VCC  25  & 115.90 &  -.21 &   9.94 \\ 
VCC  25  & 125.60 &  -.18 &   9.93 \\ 
VCC  25  & 135.20 &  -.15 &   9.93 \\ 
VCC  58  &  19.30 &  -.99 &  12.39 \\ 
VCC  58  &  29.00 &  -.82 &  11.82 \\ 
VCC  58  &  38.60 &  -.69 &  11.49 \\ 
VCC  58  &  40.60 &  -.67 &  11.44 \\ 
VCC  58  &  48.30 &  -.59 &  11.24 \\ 
VCC  58  &  51.80 &  -.56 &  11.16 \\ 
VCC  58  &  58.00 &  -.51 &  11.04 \\ 
VCC  58  &  67.60 &  -.45 &  10.90 \\ 
VCC  58  &  77.30 &  -.39 &  10.80 \\ 
VCC  58  &  86.90 &  -.34 &  10.73 \\ 
VCC  58  &  96.60 &  -.29 &  10.66 \\ 
VCC  58  & 106.30 &  -.25 &  10.61 \\ 
VCC  58  & 115.90 &  -.21 &  10.56 \\ 
VCC  58  & 125.60 &  -.18 &  10.52 \\ 
VCC  58  & 135.20 &  -.15 &  10.48 \\ 
VCC  58  & 144.90 &  -.12 &  10.46 \\ 
VCC  58  & 154.60 &  -.09 &  10.45 \\ 
VCC  58  & 164.20 &  -.06 &  10.44 \\ 
VCC  58  & 173.90 &  -.04 &  10.43 \\ 
VCC  73  &  12.90 & -1.00 &  10.60 \\ 
VCC  73  &  19.30 &  -.82 &  10.19 \\ 
VCC  73  &  25.80 &  -.70 &   9.95 \\ 
VCC  73  &  27.00 &  -.68 &   9.91 \\ 
VCC  73  &  31.60 &  -.61 &   9.81 \\ 
VCC  73  &  32.20 &  -.60 &   9.80 \\ 
VCC  73  &  35.10 &  -.56 &   9.76 \\ 
VCC  73  &  38.60 &  -.52 &   9.71 \\ 
VCC  73  &  45.10 &  -.45 &   9.64 \\ 
VCC  73  &  51.50 &  -.40 &   9.59 \\ 
VCC  73  &  58.00 &  -.35 &   9.55 \\ 
\noalign{\smallskip}
\hline
\end{array}
\]
\end{table}

\begin{figure}
\psfig{figure=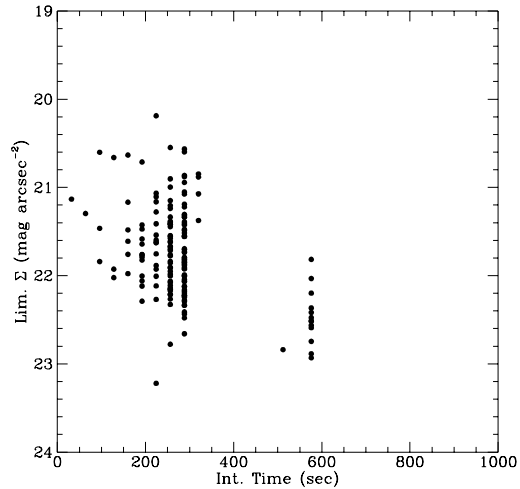,width=10cm,height=10cm}
\caption{The distribution of the limiting surface brightness reached in the outer light profiles,
as a function of the exposure time.}
\label{fig.4}
\end{figure}

\subsection{Radii $r_H(20.5)$}

The lowest 
surface brightness reached in each image is given as a function of the integration
time in Fig. 4 (see Paper III for a more comprehensive description of the
meaning of "lowest surface brightness"). Although the present observations are
deeper on average than the ones obtained at TIRGO, we decided for consistency
to measure the H band radii at the same isophotal radius as in Paper III:
i.e. at the 20.5 \marc~isophote. 

The comparison between the isophotal B band radii and the infrared $r_H(20.5)$ isophotal radii
determined in this work is shown in Fig. 5. The B radii are those 
measured on the du Pont plates at the faintest detectable 
isophote, as listed in the VCC. These are on average larger by 25\% than the standard
$r_B(25.0)$ (Binggeli et al. 1985). 
Thus it is not surprising that the relation
$r_H(20.5)~=~0.7~r_B$ used in Paper III (and reproduced as a solid line in Fig. 5)
does not hold with the present data-set.

\begin{figure}
\psfig{figure=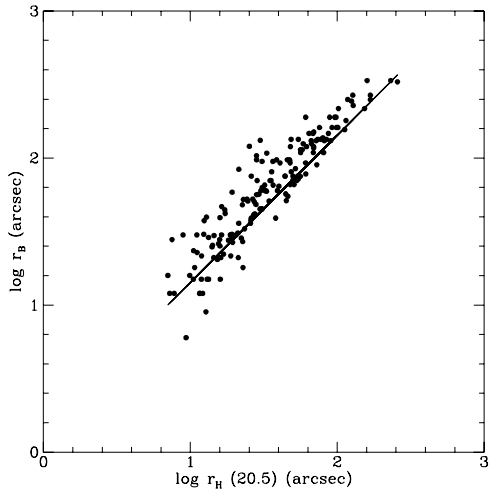,width=10cm,height=10cm}
\caption{The relation between the apparent major radius $r_H(20.5)$ 
as determined in the infrared
(this work) and the optical $r_B$ determined in the VCC at the faintest detectable 
isophote. The solid line represents the relation $r_H(20.5)~=~0.7~r_B(25.0)$.}
\label{fig.5}
\end{figure}

\subsection{Magnitudes ($H_T$, $H_{B25}$)}

$H_{B25}$ magnitudes listed in Column 15 of Table 1 are obtained by extrapolating 
the circular aperture measurements to the optical $r_B(25.0)$ radius
using standard growth curves (as in Gavazzi \& Boselli 1996). \footnote{For VCC
galaxies the optical radius is not determined at the 25 \marc isophote, but at
the faintest detectable isophote (see Section 3.2)} 
$H_T$ mag instead are obtained by extrapolating to infinity the magnitude integrated along elliptical isophotes using combinations of exponential and de Vaucoulers laws
(see Gavazzi et al. 1999: Paper V). 
As expected, $H_T$ are brighter than $H_{B25}$ by $0.05\pm0.15$~mag on average.

\subsection{Concentration index ($C_{31}$)}

The concentration index $C_{31}$ is a 
measure of the shape of light profiles in galaxies, independent of
a (model--dependent)
bulge--disk decomposition. Values larger than ($C_{31}>2.8$) indicate the presence of
substantial bulges.

We confirm the presence in our sample of a general
correlation between $C_{31}$ and the H band (total or $H_{25}$) luminosity
(computed from the redshift distance). 
We find that $C_{31}$ generally increases toward higher absolute magnitudes (Fig. 6).
High $C_{31}$ are found only among high luminosity systems, but the reverse is not true:
there are several high luminosity systems (namely late type spirals) with no or little bulge
($C_{31}\sim 3$).

\begin{figure}
\psfig{figure=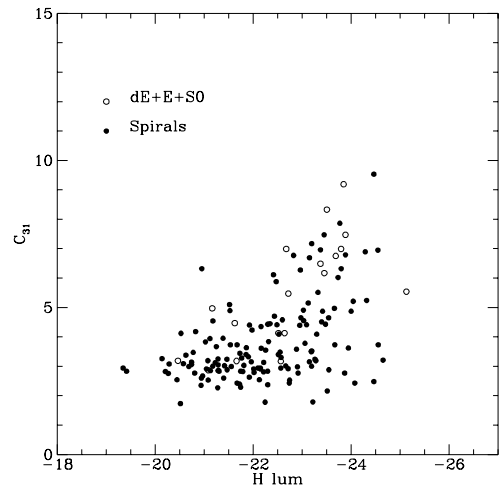,width=10cm,height=10cm}
\caption{The dependence of the near--infrared concentration index $C_{31}$ on H band
luminosity.}
\label{fig.6}
\end{figure}

\section{Summary}

We obtained images in the near-infrared H bandpass for
an optically selected ($m_p \le 15.7$) sample of 170 nearby ($z<0.02$) galaxies. 
As in previous papers we derive H magnitudes at the optical radius, total H magnitudes, isophotal
radii at the 20.5 \marc~isophote and light concentration index $C_{31}$.
As mentioned in the Introduction, the observations presented in this
paper do not cover by themselves a complete sample, but they represent the last step
of our extensive NIR survey. 
Papers I, II, III, IV of this series and B97 contain all the data gathered so far.
A comprehensive analysis of the NIR properties
of galaxies will be the subject of forthcoming papers of this series.
Paper V will report on
the profile decomposition.
\newline

\vskip 1.2cm
\acknowledgements

We wish to thank the Calar Alto Time Allocation Commetee for devoting three
nights at the 2.2m telescope for a project leaded by astronomers not belonging
to German Instituts.
We thank the MAGIC team at MPI f\"ur Astronomie for their skillful operational 
support and for several helpful discussions about data reduction. 
\newline


\vskip 1.5cm
\begin{thebibliography}{}

\bibitem[]{} Binggeli B., Sandage A.,  Tammann G., 1985, AJ, 90, 1681 
\bibitem[]{} Binggeli B., Popescu C., Tammann G., 1993, A\&AS, 98, 275
\bibitem[]{} Boselli A., Gavazzi G., 1994, A\&A, 283, 12
\bibitem[]{} Boselli A., Tuffs R., Gavazzi G., Hippelein H.,
Pierini D., 1997, A\&AS, 121, 507 (B97)
\bibitem[]{} de Vaucouleurs G., 1977, in ``Evolution of Galaxies and Stellar
Populations'', eds. R. Larson \& B. Tinsley (New Haven: Yale University
Observatory), 43
\bibitem[]{} Elias J., Frogel J., Matthews K., Neugebauer G., 1982, AJ, 87, 1029
\bibitem[]{} Gavazzi G., Boselli A., 1996, Ap.Lett.\& Comm, 35, 1
\bibitem[]{} Gavazzi G., Pierini D., Boselli A., Tuffs R., 1996c, A\&AS,
120, 489 (Paper I)
\bibitem[]{} Gavazzi G., Pierini D., Baffa C., Lisi F., Hunt L., Boselli, A., 1996b, A\&AS,
120, 521 (Paper II)
\bibitem[]{} Gavazzi G., Franzetti P., Scodeggio M., et al., 1999, A\&AS, this issue (Paper III)
\bibitem[]{} Gavazzi G., Franzetti P., Scodeggio M., Boselli A., Pierini D.,  1999a, A\&AS (submitted) (Paper V)
\bibitem[]{} Gezari D., Schmitz M., Pitts P., Mead J., 1993, Nasa Reference Publication, 1294
\bibitem[]{} Herbst T., Beckwith S., Birk C., et al, 1993, SPIE 1946
\bibitem[]{} Wainscoat R., Cowie L. 1992, AJ, 103, 332
\bibitem[]{} Zwicky F., Herzog E., Karpowicz M., Kowal C., Wild P., 
1961-1968, ``Catalogue of Galaxies and Clusters of Galaxies'', 6 vol., Pasadena, C.I.T.
\end{thebibliography}
\end{document}